\documentclass[times, twoside]{zHenriquesLab-StyleBioRxiv}

\usepackage{amsmath,amsfonts,amssymb}
\usepackage{booktabs}
\usepackage{multirow}
\leadauthor{Jansen} 

\begin{document}

\title{Patient-specific fine-tuning of CNNs for follow-up lesion quantification}
\shorttitle{Patient-specific fine-tuning of CNNs}

\author[1,\Letter]{Mari\"elle J.A. Jansen}
\author[1,]{Hugo J. Kuijf}
\author[2,]{Ashis K. Dhara}
\author[3,]{Nick A. Weaver}
\author[3,]{Geert Jan Biessels}
\author[2,]{Robin Strand}
\author[1,]{Josien P.W. Pluim}

\affil[1]{UMC Utrecht and Utrecht University, Image Sciences Institute, P.O. Box 85500, Q02.445, 3508 GA Utrecht, the Netherlands}
\affil[2]{Center for Image Analysis, Department of Technology, Uppsala University, Box 337, SE-75105 Uppsala, Sweden}
\affil[3]{UMC Utrecht, Department of Neurology, Brain Center Rudolf Magnus, P.O. Box 85500, G03.232, 3508 GA Utrecht, the Netherlands}

\maketitle

\begin{abstract}
Convolutional neural network (CNN) methods have been proposed to quantify lesions in medical imaging. Commonly more than one imaging examination is available for a patient, but the serial information in these images often remains unused. CNN-based methods have the potential to extract valuable information from previously acquired imaging to better quantify current imaging of the same patient.\par
A pre-trained CNN can be updated with a patient’s previously acquired imaging: patient-specific fine-tuning. In this work, we studied the improvement in performance of lesion quantification methods on MR images after fine-tuning compared to a base CNN. We applied the method to two different approaches: the detection of liver metastases and the segmentation of brain white matter hyperintensities (WMH).\par
The patient-specific fine-tuned CNN has a better performance than the base CNN. For the liver metastases, the median true positive rate increases from 0.67 to 0.85. For the WMH segmentation, the mean Dice similarity coefficient increases from 0.82 to 0.87. In this study we showed that patient-specific fine-tuning has potential to improve the lesion quantification performance of general CNNs by exploiting the patient’s previously acquired imaging. 

\end {abstract}

\begin{corrauthor}
marielle\at isi.uu.nl
\end{corrauthor}

\section*{Introduction}
Lesion quantification is an important step in medical image analysis. Over the years convolutional neural network (CNN) based lesion quantification methods have been proposed for medical images\textsuperscript{1–5} to aid radiologists in the detection and segmentation of these lesions. For some diseases or treatments, there is a clinical need to follow-up on the patient. For these patients more than one imaging examination is available. The previously acquired scan, i.e. baseline scan, contains patient-specific information that could be exploited by the CNN-based method to better quantify the current scan of the same patient, i.e. the follow-up scan.\par
The conventional way to train a CNN is to iterate over a large data set representing the object to detect or segment. In this way, a method is obtained that should give reliable results in comparable data sets. However, CNN-based methods do not always provide satisfying performance for clinical practice. Differences in image quality and variations among patients are often not fully covered within the method and can cause insufficient results. As a result of training a CNN on a general population, the CNN is not fully adapted to the specific details of an individual patient. \par
Updating a CNN toward the features of a patient could improve the outcome of the CNN for that specific patient\textsuperscript{6,7}. Previously acquired scans can be used for this purpose, as they are already analyzed at the time of examination of the follow-up scan. We therefore propose a method to enhance the performance of lesion quantification in a follow-up scan with a patient-specific fine-tuning approach. In addition, we gained insight on the necessary conditions to achieve these improvements.\par
Fine-tuning a pre-trained CNN has the benefit of obtaining good results using only a small data set during the fine-tuning step and has successfully been applied in previous studies. For example, medical image domain knowledge has been transferred to a CNN pre-trained on natural images by fine-tuning the CNN with medical images\textsuperscript{8,9}. Pre-trained CNNs have been fine-tuned towards the features of one specific image, resulting in better segmentations of that image\textsuperscript{10,11}.
Patient-specific fine-tuning of a CNN is a method to update the CNN using a previous scan of a patient. This fine-tuned CNN has learned the specific features of abnormalities and the surrounding healthy tissue to improve the detection or segmentation in a follow-up scan. This approach is based on the assumption that the baseline and follow-up scan of a patient share the features of healthy tissue and abnormalities. \par
We present and evaluate the patient-specific fine-tuning approach on two applications; the detection of liver metastases on MRI and the segmentation of brain white matter hyperintensities (WMH) on MRI. Furthermore, different aspects of patient-specific fine-tuning are studied.\par

\section*{Materials and methods}
The proposed lesion quantification framework is shown in Figure 1. First a CNN is trained using the training set, referred to as the base CNN model. Next, this base CNN is refined for each individual patient in the patient-specific fine-tuning step. A previous MRI scan of a patient, referred to as baseline scan, is used to fine-tune the network to the specific features of that patient and its lesions. During the testing step this patient-specific CNN model is used to detect or segment lesions in a follow-up MRI scan of the same patient. Two MR sequences are applied to train and test the CNN model. \par

\begin{figure}
	\begin{center}
		\begin{tabular}{c} 
			\includegraphics[width=8.0cm]{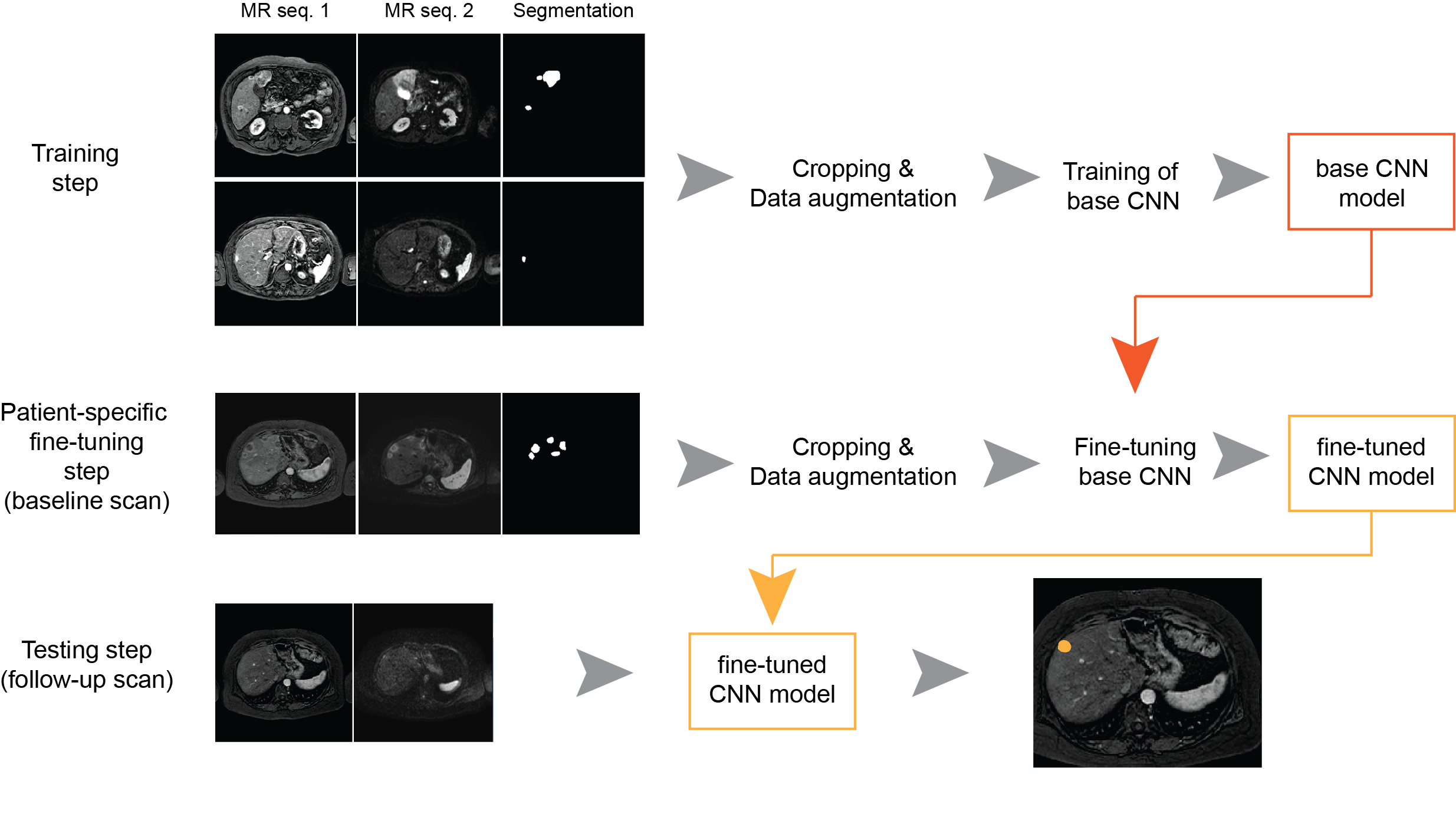}
		\end{tabular}
	\end{center}
	\caption[example]
	{ \label{fig:Fig1} 
		 Proposed lesion quantification framework, shown with the liver MRI. First a base CNN is trained with a training set consisting of multiple patients. Next, the base CNN is refined in the patient-specific fine-tuning step using a previous MRI exam of a patient (the baseline scan). The fine-tuned CNN is used to detect or segment lesions in a follow-up MRI scan of the same patient. The images are cropped to focus of the organ of interest. The cropped image size is 128$\times$128.}
\end{figure}

\subsection*{Data}
The patient-specific fine-tuning approach is demonstrated on two different data sets: abdominal MRI for liver metastases detection and brain MRI for WMH segmentation. The UMCU Medical Ethical Committee has reviewed and approved this study and informed consent was waived due to its retrospective nature.\par
\subsubsection*{Liver metastases detection}
In this study, abdominal MRI of 47 patients with liver metastases from the University Medical Center Utrecht, the Netherlands, were included. Sixteen of them had at least two consecutive MRI examinations within a three months to a year. All patients underwent a clinical MRI examination, including dynamic contrast enhanced (DCE) MR series and diffusion weighted (DW) MRI, acquired on a 1.5T scanner (Philips, Best, the Netherlands). These two MR sequences are used to train the CNN.\par
The DCE-MR series was acquired in six breath holds with one to five 3D images per breath hold, with in total 16 3D images. The DW-MRI was acquired with three b-values: 10, 150, and 1000 s/mm2. The DCE-MRI was corrected for motion using a principle component analysis (PCA) based groupwise registration\textsuperscript{12}. Intensity normalization was applied to both MR sequences and the DW-MRI was registered to the DCE-MRI by a rigid transformation, followed by a b-spline transformation (The parameter files used, are available at elastix.bigr.nl/wiki/index.php/Par0057). The resulting images have 100 slices and matrix sizes of 256 $\times$ 256. Voxel size is 1.543 mm $\times$ 1.543 mm $\times$ 2 mm.\par
The liver metastases were manually segmented on the DCE-MRI by a radiologist in training and verified by a radiologist with more than ten years of experience. The data set included mainly colorectal metastases, neuroendocrine metastases, and some other metastasis types (i.e. other gastrointestinal metastases and breast metastases). On average 30\% of the liver slices contained liver metastases. Liver masks were automatically obtained using our previously developed segmentation method\textsuperscript{13}. \par
MRI data from 31 patients of which only one MRI examination was available were used to train the base CNN. The slices for training were limited to the slices containing metastases for a more balanced data set, resulting in a total number of 798 2D slices for training. The remaining 16 patients, with available baseline and follow-up scans, were used for testing. The scans had an average of 6 metastases per patient, ranging from 1 to 31 metastases.\par

\subsubsection*{WMH segmentation}
The brain MRI data of 80 memory clinic patients with WMH were included in the study. Twenty of the patients were from the Dutch Parelsnoer Institute - neurodegenerative diseases study\textsuperscript{14}, from the UMC Utrecht, the Netherlands, and the remaining 60 were from the WMH challenge training set\textsuperscript{15}. The 20 patients from UMC Utrecht had two MRI examinations, with two years between the two examinations. \par
All MR exams were acquired with a similar protocol, with a T2-weighted-Fluid-Attenuated Inversion Recovery (FLAIR) MRI and a T1-weighted MRI, acquired on a 3T scanner16. These two MR sequences are used to train the CNN. Both MR sequences were bias field corrected and the T1-weighted MRI were registered to the FLAIR images; more details can be found in Kuijf et al.\textsuperscript{15}. All images were resized to a matrix size of 240 $\times$ 240 and 48 slices. Voxel size is 0.958 mm $\times$ 0.958 mm $\times$ 3.00 mm. \par
The WMH were manually segmented on the FLAIR images by an experienced researcher, in accordance with the STRIVE criteria\textsuperscript{17}. WMH segmentation was performed with in-house developed software on MeVisLab (MeVis Medical Solutions AG, Bremen, Germany). On average 63\% of the brain slices contained WMH. Brain masks were obtained using the SPM software\textsuperscript{18}.
The MRI data of 60 patients with a single MRI examination from the WMH challenge were used for training. The slices for training were limited to the slices containing WMH lesions for a more balanced data set, resulting in a total of 1383 2D slices for training the base CNN. The remaining 20 patients, with a baseline and follow-up scan, were used for testing. These scans had an average of 65 WMH per patient, ranging from 21 to 117 WMH.\par

\subsection*{Base CNN model}
The overall architecture of the CNN was inspired by our earlier work on liver metastases detection\textsuperscript{13}. This fully convolutional architecture includes elements from the P-net architecture, which has proven to be efficient for updating\textsuperscript{10}. As most MRI exams usually consist of multiple MR sequences, we modified the original P-net to include a dual pathway that can process two MR sequences, each in a separate pathway to extract specific feature maps for those sequences. The input image for each pathway was one MR sequence. If the MR sequence had multiple instances, such as the phases of the abdominal DCE-MRI, the 2D images were combined into one input image with the instances as channels. Variations on the network architecture have been tested in our earlier work\textsuperscript{13}. See Figure 2 for an overview of the fully convolutional network architecture. \par

\begin{figure}
	\begin{center}
		\begin{tabular}{c} 
			\includegraphics[width=8.0cm]{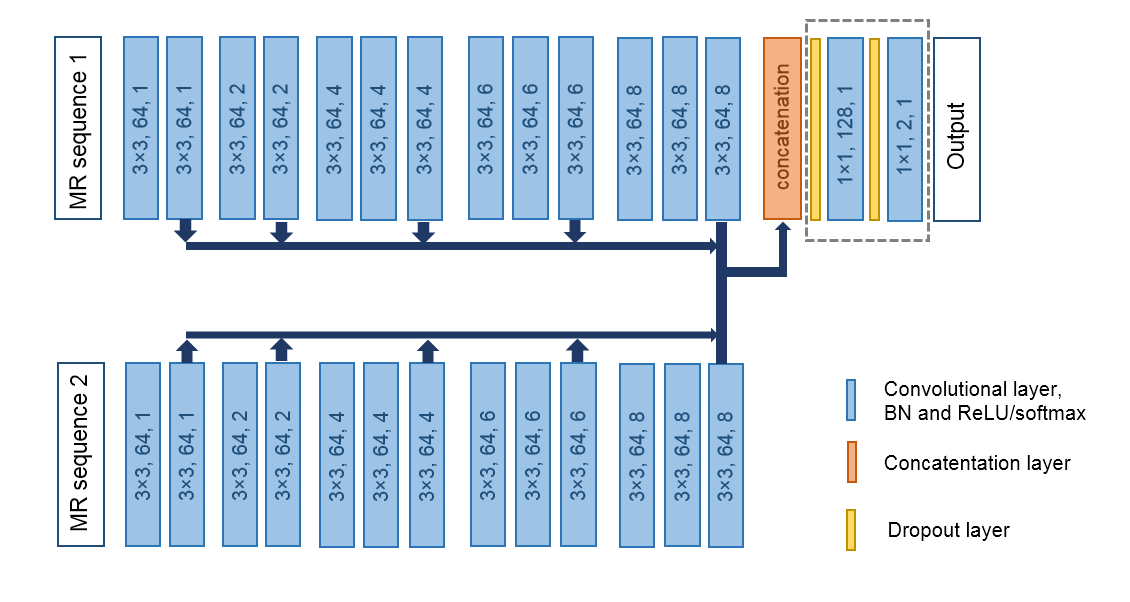}
		\end{tabular}
	\end{center}
	\caption[example]
	{ \label{fig:Fig2} 
		 Fully convolutional network architecture for lesion quantification. The blue blocks represent the convolutional layers, batch normalization (BN) and ReLU or softmax activation. The size of the kernel of each convolutional layer is given in the block, followed by the number of kernels and the dilation rate of the kernel. The dashed line indicates the trainable layers during the patient-specific fine-tuning step.}
\end{figure}

Each pathway had 13 convolutional layers (2 convolutional layers and 11 dilated convolutional layers\textsuperscript{19} with varying dilation rates), each having 3$\times$3 convolution and 64 kernels, split in five blocks. The feature maps at the end of each block of each pathway were concatenated in the third dimension, resulting in a feature map with 640 kernels, and were passed to two convolutional layers with a 1$\times$1 convolution with 128 and 2 kernels, respectively. This resulted in a receptive field of 123$\times$123 pixels. \par
During training, categorical cross-entropy was used as the loss function. This loss function was calculated on a pixel level resulting in detection by segmentation. ReLU activation and batch normalization were used in all the convolutional layers, except for the final layer, which had a softmax activation. Two dropout layers were applied before and after the second-last layer. The dropout rate was set to 0.2. The classes were weighted based on class frequencies. The class weights were set to 1 for the background class and to 5 for the lesion class. He uniform was used as initializer and Adam as optimizer with a learning rate of 0.0001. The network was trained for 10,000 iterations, with 4 images per mini batch. 
Twenty-five patches of 128$\times$128 pixels were taken from each slice for data augmentation. The patches originated from the organ of interest and have overlapping areas. Online data augmentation was applied by random rotation of the patches, with rotation angles of $\pm$45 degrees.

\subsection*{Patient-specific fien-tuning}
In the patient-specific fine-tuning step the last two layers of the base CNN were further trained with the baseline scan of a patient, to improve the lesion quantification results in the follow-up scan of the same patient. Only the last two layers were updated, as these combine the feature maps of the previous blocks of the two pathways and to ensure a fast fine-tuning step.\par
The CNN was refined by continuing training of the base CNN with the weights of all layers frozen, i.e. untrainable, except for those of the last two layers. The two trainable layers are indicated with the dashed line box in Figure 2. \par
The same loss function, optimizer, and learning rate were used as the ones during the training of the base CNN. The CNN was refined with 4 images per mini batch and the number of iterations was optimized during the experiments, see the Experiments section. The images used for fine-tuning were MRI slices of the patient’s baseline scan. The baseline scan was annotated earlier by an expert and only slices with at least one lesion present were included in the fine-tuning step. Five patches of 128$\times$128 pixels were taken within the organ of interest region from each slice, four patches from the corners of the region and one center patch, removing air and tissues not of interest from the input images. The patches were randomly rotated, with rotation angles of $\pm$45 degrees. Rotation of the patches included the variance in the positioning of the body during the MRI examination. \par

\subsection*{Lesion quantification on follow-up scan}
The follow-up data was processed by the individual patient-specific CNN, i.e. the base CNN fine-tuned with the baseline scan of that patient. The probability output of the network was masked by the liver or brain mask.\par
For the evaluation, the masked probability output was post-processed to a binary image. A threshold of 0.5 was applied to the softmax output of the network and morphological closing with a structuring element of 3$\times$3$\times$3 was applied to fill holes. For the liver data set, the morphological closing was followed by a morphological opening with a plus-shaped structuring element of 3$\times$3, to remove noise pixels. This was not applied to the brain data set as WMHFF of 1 pixel occur frequently. For the liver data set, the resulting binary image was divided into separate objects representing individual lesions, using voxel clustering with 26-neighbourhood connection.

\section*{Experiments}
The setup for the base CNN model and the process of fine-tuning the CNN, as explained above, are similar for both the liver metastases detection and the WMH segmentation. The same experiments are conducted, only the evaluation is specified for the task: detection or segmentation. \par 

\subsection*{Liver metastases detection}
The detection of (new) liver metastases is important to monitor disease progression and for treatment planning. Treatment selection is based on the detection findings, i.e. the location and the number of metastases. Progress is monitored by follow-up scans.\par
The detection efficacy is evaluated using the true positive rate (TPR), the number of false positives per case (FPC), and the F1 score. As the values of the TPR, FPC, and F1 score do not have a normal distribution, the median and interquartile range (IQR) are reported and the Wilcoxon signed rank test will be used to test for significant differences.\par
The TPR is calculated as the number of true positive objects divided by the total number of true lesion objects. A lesion is considered detected, and thus a true positive object, when the manual annotation and the predicted segmentations have an overlap greater than 0. The FPC is calculated as the number of detected objects not overlapping with any true lesion object. The F1 score is calculated as $(2*recall*precision)/(recall+precision)$, where recall is the TPR as defined above and precision is the number of true positive objects divided by the total number of detected objects (both true positive and false positive objects).  \par

\subsection*{WMH segmentation}
WMH are a common radiological finding in the elderly population, and are generally considered to reflect cerebral small vessel disease.\textsuperscript{20} The presence and extent of WMH are associated with cognitive decline and dementia.\textsuperscript{17,20,21} In particular, progression of WMH over time has been linked to cognitive decline and risk of dementia.\textsuperscript{22,23} Quantification of WMH volume changes over time could contribute to the monitoring of disease progression, and provide clinicians with relevant information for informed decisions.\par
WMH segmentation on brain MRI is evaluated per MRI exam using the overall Dice Similarity Coefficient (DSC) and the absolute volume difference (AVD). The Dice score is calculated as $(2*X\cap Y)/(X+Y)$ , where $X$ is the automatic segmentation and $Y$ the manual annotation. The AVD is calculated as $(abs(V(X)-V(Y)))/V(Y) *100\%$, where $V(X)$ is the automatic segmentation volume and $V(Y)$ the manual annotation volume. The mean and standard deviation are reported and the paired Student’s t-test is used to test for significant differences.\par
The following experiments study different elements of patient-specific fine-tuning; the number of iterations, the number of slices presented during fine-tuning, and a weighting scheme.

\subsection*{Number of iterations}
The duration of the fine-tuning should be adapted to the similarity between the baseline and the follow-up scan, as the baseline scan can have a different appearance to the follow-up scan. For example, liver metastases show changes in shape and size after treatment, and WMH are known to progress from punctuate to confluent lesions over time. If the CNN is fine-tuned for too few iterations the CNN will not be patient specific enough. If the number of iterations is too high, the risk of overfitting towards the baseline scan increases. The number of iterations is therefore studied and the optimal number will be used in further experiments. The number of iterations explored ranges from 50 to 1000. \par

\subsection*{Number of slices}
The possibility of fine-tuning with only one or two slices is studied. Annotating or correcting the annotation of the baseline scan on one or two slices would be a smaller effort than annotating a full image if a good annotation does not yet exist. In these experiments, the number of slices that are offered to the CNN are one, two, or all slices with lesions or the full organ of interest present. \par
When not all slices are included, selection is based on the softmax probability outcome of the base CNN on the baseline scan. The slices with an average softmax probability closest to 0.5 are selected, assuming that these slices contain the most valuable information for updating the CNN. Two options were explored for the slice selection procedure. The first is including all slices for slice selection and the second option is only including slices with lesions present for slice selection.\par

\subsection*{Weighting}
Weighting false negatives, false positives, and true positives, obtained after performing the lesion quantification with the base CNN on the baseline scan, will redefine what we want the CNN to learn during fine-tuning. The weights are set according the task to be performed. \par
For the detection task, the objective is to detect the metastases on liver MRI. The weights of the missed metastases are set to the highest value (five). The weights of the true positive pixels in detected metastases are set to two, with the weights of false negatives and the false positive pixels connected to the detected metastases set to zero. The background and false positive objects weights are set to one. In this manner the fine-tuned CNN will be greedier in labelling pixels as liver metastasis. Over- or undersegmentation of detected metastases is not considered incorrect labelling and false positive objects are considered less problematic than missing a metastasis. \par
For the segmentation task, the objective is to segment the WMH on brain MRI and get a good lesion volume estimation. The weights of the false positive and false negative pixels are set to the highest value (five), as the main goal is to reduce the incorrectly labelled pixels. The weights of the true positive pixels are set to two, and the true negative pixels are set to one, to handle class imbalance. \par
The range of values of the weights are based on the class weights during training of the base CNN, which were five and one for lesions and background, respectively.\par

\subsection*{Uncertainty of CNN}
The uncertainty of the CNN in detecting or segmenting the lesions is assessed by the standard deviation (SD) in probability outcome of the softmax layer. We implemented Monte Carlo dropout during test time\textsuperscript{24,25}, repeating the test phase 25 times. The SD of the probabilities over the 25 repetitions is calculated for every voxel. For the detection task the mean SD of the detected metastases is calculated as uncertainty metric, to study the change in network uncertainty for detecting metastases. For the segmentation task the maximum SD of all voxels in an image is taken as uncertainty metric, to study the change in network uncertainty for the entire image.\par 

\section*{Results}

\subsection*{Liver metastases detection}
\subsubsection*{Number of iterations}
In Table 1 the TPR, FPC, and the F1 score are reported for different numbers of iterations during the fine-tuning step. For a duration of only 50 iterations the TPR is similar to the TPR of the base CNN, but for 100 iterations and higher the TPR improves. However, for more than 50 iterations the FPC increases slightly. No significant differences (Wilcoxon signed rank test with p<0.01) are found between the base CNN and the fine-tuned CNN for different number of iterations. For the remaining experiments, the CNN is fine-tuned for 100 iterations.

\begin{table}
    \centering
    \caption{Median [interquartile range] of the true positive rate (TPR), the false positive count (FPC) and the F1 score of the liver metastases detection, for a varying number of iterations (iter.) of learning for the CNN for fine-tuning. The best results are printed in bold.}
	\label{Table1}
    \begin{tabular}{llll}
        \toprule
                & \textbf{TPR}                  & \textbf{FPC}         & \textbf{F1 Score}             \\
                \midrule
                
Base CNN        & 0.67 {[}0.32-1.00{]}          & 3 {[}0-7{]}          & 0.49 {[}0.21-0.67{]}          \\
50 iter.   & 0.67 {[}0.20-1.00{]}          & \textbf{1 {[}0-3{]}} & 0.40 {[}0.25-0.80{]}          \\
100 iter.  & 0.85 {[}0.44-1.00{]}          & 2 {[}0-3{]}          & \textbf{0.57 {[}0.40-0.67{]}} \\
500 iter.  & \textbf{0.92 {[}0.33-1.00{]}} & 2 {[}0-4{]}          & 0.50 {[}0.25-0.71{]}          \\
1000 iter. & 0.85 {[}0.50-1.00{]}          & 1 {[}0-4{]}          & 0.50 {[}0.40-0.75{]}\\     
    \bottomrule    
    \end{tabular}
\end{table}

\subsubsection*{Number of slices}
The median [IQR] of the TPR, FPC, and the F1 score for the inclusion of different (numbers of) slices are presented in Table 2. The slice selection considered either all liver slices or all slices containing metastases. The slices with an average softmax probability closest to 0.5 were selected.
Including all the slices with liver metastases for fine-tuning the CNN gives the best results for the liver metastases detection method. Fine-tuning the CNN with only one or two selected slices of the liver does not improve the results.

\begin{table}
\centering
    \caption{Median [interquartile range] of the true positive rate (TPR), the false positive count (FPC) and the F1 score for a ranging number of slices presented to the CNN for fine-tuning. The best results are printed in bold. No significant differences were found between the Base CNN and all options.}
	\label{Table2}
\begin{tabular}{llll}
\toprule
                  & \textbf{TPR}                  & \textbf{FPC}         & \textbf{F1 Score}             \\ \midrule
Base CNN          & 0.67 {[}0.32-1.00{]}          & 3 {[}0-7{]}          & 0.49 {[}0.21-0.67{]}          \\
1 slice liver     & 0.50 {[}0.11-1.00{]}          & 4 {[}1-8{]}          & 0.33 {[}0.12-0.64{]}          \\
2 slices liver    & 0.42 {[}0.11-1.00{]}          & 1 {[}0-5{]}          & 0.33 {[}0.15-0.67{]}          \\
1 slice metas.    & 0.67 {[}0.08-1.00{]}          & 4 {[}2-6{]}          & 0.31 {[}0.00-0.50{]}          \\
2 slices metas.   & 0.67 {[}0.08-1.00{]}          & 3 {[}1-5{]}          & 0.31 {[}0.00-0.44{]}          \\
\begin{tabular}[c]{@{}l@{}}All metas. \\  slices\end{tabular} & \textbf{0.85 {[}0.44-1.00{]}} & 2 {[}0-3{]}          & \textbf{0.57 {[}0.40-0.67{]}} \\
All liver slices  & 0.67 {[}0.20-1.00{]}          & \textbf{0 {[}0-3{]}} & 0.56 {[}0.33-0.80{]}          \\ \bottomrule
\end{tabular}
\end{table}

\subsubsection*{Weighting}
The median [IQR] of the TPR, FPC and F1 score for the base CNN, fine-tuned CNN without weights and fine-tuned CNN with weights are presented in Table 3. The fine-tuning was done for 100 iterations and included all metastases slices. Putting more weight on the missed liver metastases, leads to more detected metastases, but at the cost of more false positive objects.

\begin{table}
\centering
    \caption{Median [interquartile range] of the true positive rate (TPR), the false positive count (FPC) and the F1 score of the liver metastases detection, for weighting the true positives, false negatives and false positives during the patient-specific fine-tuning (FT). The best results are printed in bold. An asterix indicates a significant difference with results of the ‘Base CNN’ (Wilcoxon signed rank test with p<0.01).}
	\label{Table3}
\begin{tabular}{llll}
\toprule
                  & \textbf{TPR}                  & \textbf{FPC}         & \textbf{F1 Score}             \\ \midrule
Base CNN          & 0.67 {[}0.32-1.00{]}          & 3 {[}0-7{]}          & 0.49 {[}0.21-0.67{]}          \\
FT CNN & 0.85 {[}0.44-1.00{]} & \textbf{2 {[}0-3{]}}         & \textbf{0.57 {[}0.40-0.67{]}} \\
\begin{tabular}[c]{@{}l@{}}Weighted \\  FT CNN\end{tabular}  & \textbf{1.00 {[}0.62-1.00{]}}          & 6 {[}3-14{]}\textsuperscript{*} & 0.42 {[}0.29-0.57{]}          \\ \bottomrule
\end{tabular}
\end{table}

\subsubsection*{Qualitative results}
Some visual examples of lesion detection by the base CNN and by the patient-specific CNN are given in Figure 3. The patient-specific CNN is fine-tuned in 100 iterations, including all slices with metastases present, and without weights. After patient-specific fine-tuning the median TPR increases from 0.67 to 0.85, the median FPC decreases from 3 to 2, and the F1 score increases from 0.49 to 0.57. \par

\begin{figure}
	\begin{center}
		\begin{tabular}{c} 
			\includegraphics[width=8.2cm]{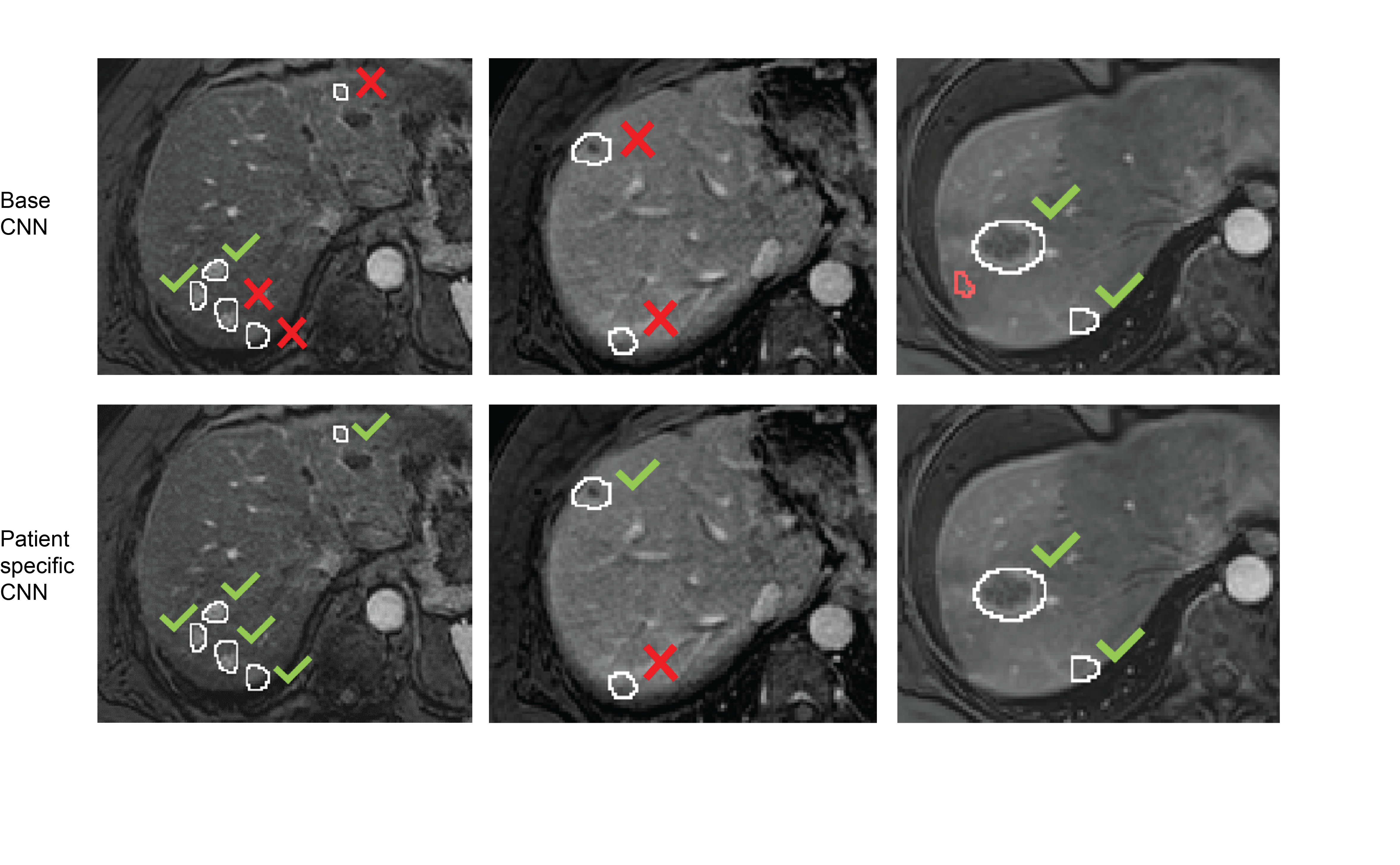}
		\end{tabular}
	\end{center}
	\caption[example]
	{ \label{fig:Fig3} 
		 Examples of the detection results of the base CNN and the patient-specific CNN for three different patients. White outline = manual annotation, red outline = false positive object, green check = detected metastasis, red cross = missed metastasis.}
\end{figure}

Metastases smaller than 1.0 cm³ are more often missed by the base CNN than larger metastases, an example of this can be seen in the first column of Figure 3. The TPR for the small metastases increases from 0.13 to 0.26 after patient-specific fine-tuning, while the TPR of the large metastases only increases slightly, from 0.81 to 0.83. The increase in TPR is thus mainly due to the higher TPR of small metastases.\par

\subsubsection*{Uncertainty of CNN}
The softmax probabilities of the detected metastases by the base CNN have a mean SD of 0.203 ($\pm$0.078). The mean SD can be considered a measure of the uncertainty of detection.  The patient-specific CNN has a mean of 0.158 ($\pm$0.070), which is significantly lower than the mean SDs of the base CNN (p=0.003, paired Student’s t-test). The patient-specific CNN does not only detect more metastases, it is also more certain about the detected metastases.\par

\subsection*{WMH segmentation}

\subsubsection*{Number of iterations}
All four options (50, 100, 500, or 100 iterations) for the duration of the fine-tuning, give similar results. The Dice score significantly increases (Paired Student’s t-test with p<0.01) and the AVD decreases after fine-tuning in comparison with the results of the base CNN. The base CNN has a mean DSC of 0.82 and a mean AVD of 20.7\%. The fine-tuned CNNs have a mean DSC around 0.87 and a mean AVD around 10\%. Using only 50 iterations gives a slightly lower Dice score (0.85) than the higher number of iterations. The number of iterations is set to 100 for the rest of the experiments.\par

\subsubsection*{Number of slices}
The mean ($\pm$ SD) of the Dice score and the AVD for different numbers of slices are presented in Table 4. The slice selection method selected the same slices when either all brain slices or all WMH slices were considered for selection. The selected slices originated mostly from the area around the ventricles and the regions with large WMH.
Including one slice already alters the Dice score and the AVD significantly. Meanwhile, further inclusion of all lesion or brain slices gives only a slightly better result. It is therefore possible to only annotate the one or two selected slices and improve the segmentation using these slices in the fine-tuning process.

\begin{table}
\centering
    \caption{Mean ($\pm$SD) of the Dice score and absolute volume difference (AVD) of the WMH segmentation for a varying number of slices for fine-tuning. The best results are printed in bold. A cross indicates a significant difference with results of the ‘Base CNN’ (Paired Student’s t-test with p<0.01).}
	\label{Table4}
\begin{tabular}{lll}
\toprule
                  & \multicolumn{1}{c}{\textbf{Dice score}} & \multicolumn{1}{c}{\textbf{AVD (\%)}} \\ \midrule
Base CNN          & 0.82 $\pm$ 0.05                         & 20.7 $\pm$ 13.5                       \\
1 slice           & 0.86 $\pm$ 0.04\textsuperscript{$\dagger$}                        & 11.2 $\pm$ 8.2                        \\
2 slices          & 0.86 $\pm$ 0.04\textsuperscript{$\dagger$}                          & 10.7 $\pm$ 7.9                        \\
All lesion slices & \textbf{0.87 $\pm$ 0.04}\textsuperscript{$\dagger$}                 & 10.7 $\pm$ 7.3                        \\
All brain slices  & \textbf{0.87 $\pm$ 0.04}\textsuperscript{$\dagger$}               & \textbf{10.2 $\pm$ 6.9}\textsuperscript{$\dagger$}                \\ \bottomrule
\end{tabular}
\end{table}

\subsubsection*{Weighting}
In Table 5 the mean ($\pm$ SD) of the Dice score and the AVD are presented for the base CNN, fine-tuned CNN without weights and fine-tuned CNN with weights. Visual inspection of the results showed that weighting the true positive, false positive, and false negative pixels makes the CNN more prone to label a pixel as lesion, resulting in oversegmentation. For the WMH lesion segmentation task weighting the pixels did not result in a better segmentation.\par
\begin{table}[]
\centering
    \caption{Mean ($\pm$SD) of the Dice score and absolute volume difference (AVD) of the WMH segmentation for weighting the true positives, false negatives and false positives during the patient-specific fine-tuning (FT). The best results are printed in bold. A cross indicates a significant difference with results of the ‘Base CNN’ (Paired Student’s t-test with p<0.01).}
	\label{Table5}
\begin{tabular}{lll}
\toprule
                  & \multicolumn{1}{c}{\textbf{Dice score}} & \multicolumn{1}{c}{\textbf{AVD (\%)}} \\ \midrule
Base CNN          & 0.82 $\pm$ 0.05                         & 20.7 $\pm$ 13.5                       \\

FT CNN & \textbf{0.87 $\pm$ 0.04}\textsuperscript{$\dagger$}                 & \textbf{10.7 $\pm$ 7.3}                        \\
Weighted FT CNN  & 0.86 $\pm$ 0.05\textsuperscript{$\dagger$}               & 11.8 $\pm$ 10.5              \\ \bottomrule
\end{tabular}
\end{table}

\subsubsection*{Qualitative results}
Some visual examples of the lesion segmentation on the follow-up scan by the base CNN and the patient-specific CNN are given in Figure 4. The patient-specific CNN is fine-tuned in 100 iterations, including all slices with WMH lesions present, and without weights. After patient-specific fine-tuning the average Dice score increases from 0.82 to 0.87, the AVD decreases from 20.7\% to 10.7\%. \par
Other metrics of evaluation in the WMH challenge are the modified Hausdorff distance, TPR, and F1 score\textsuperscript{15}. After patient-specific fine-tuning the Hausdorff distance decreases from 2.40 mm to 2.01 mm, the TPR decreases from 0.84 to 0.63, and the F1 score remains 0.72. The TPR decreases, because the fine-tuned CNN misses lesions with the size of only a few pixels. However, the other metrics improve due to the decrease in false positives and better volume segmentation.\par
\begin{figure}
	\begin{center}
		\begin{tabular}{c} 
			\includegraphics[width=8.2cm]{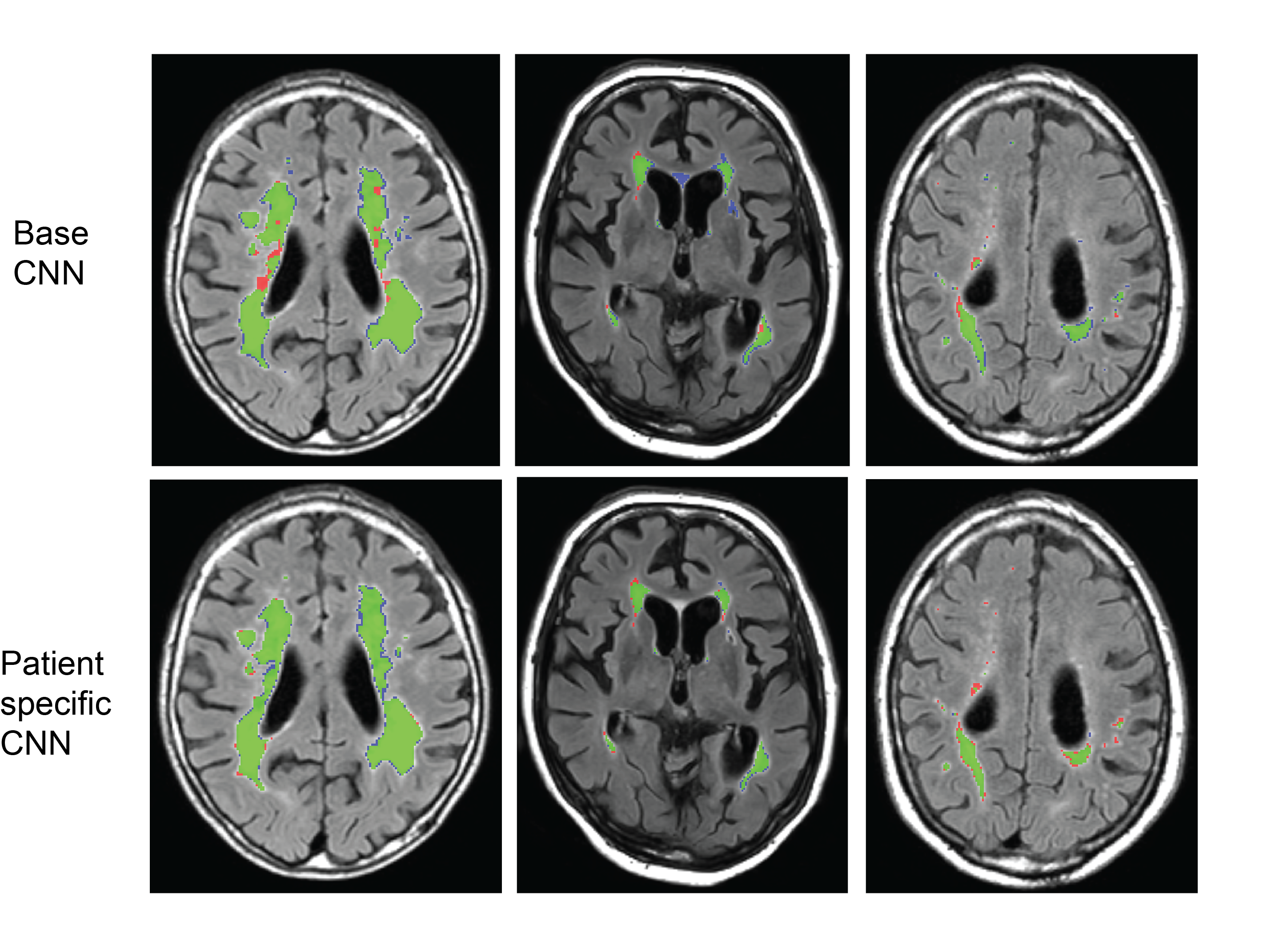}
		\end{tabular}
	\end{center}
	\caption[example]
	{ \label{fig:Fig4} 
		 Examples of the follow-up scan with the segmentation results of the base CNN and the patient-specific CNN for three different patients. Green = true positive pixels, red = false negative pixels, and blue = false positive pixels.}
\end{figure}

The majority of the incorrectly labelled pixels by the base CNN are either small false positive regions (e.g. see Figure 4, second column) or false negative pixels as part of larger lesions (e.g. see Figure 4, first column). The patient-specific CNN learns to label these pixels correctly, resulting in a higher Dice score and lower AVD and thus providing a better WMH segmentation.\par
Smaller lesions are harder to segment than larger lesions15 and we noticed that the patient-specific CNN missed more smaller lesions (<0.01 cm³) than the base CNN in the segmentation, but at the same time also segmented fewer false positive pixels. The lesions and (noisy) false positive pixels labelled by the base CNN have a similar appearance, resulting in either labelling them all as lesion or all as background. The fine-tuning procedure seems to put more emphasis on reducing the smallest false positives, at the cost of small false negatives. Figure 4, column 3 shows an example of these mislabeled pixels representing small lesions.

\subsubsection*{Uncertainty of CNN}
The softmax probabilities of the base CNN have a mean maximum SD of 0.398 ($\pm$0.025). The patient-specific CNN has a mean maximum SD of 0.328 ($\pm$0.038), which is significantly lower than the maximum SDs of the base CNN (p<0.001, paired Student’s t-test). In Figure 5, an example is given of the mapped SD of the probabilities. The patient-specific CNN is more certain about the labels given, especially within large WMH.
\begin{figure}
	\begin{center}
		\begin{tabular}{c} 
			\includegraphics[width=8.2cm]{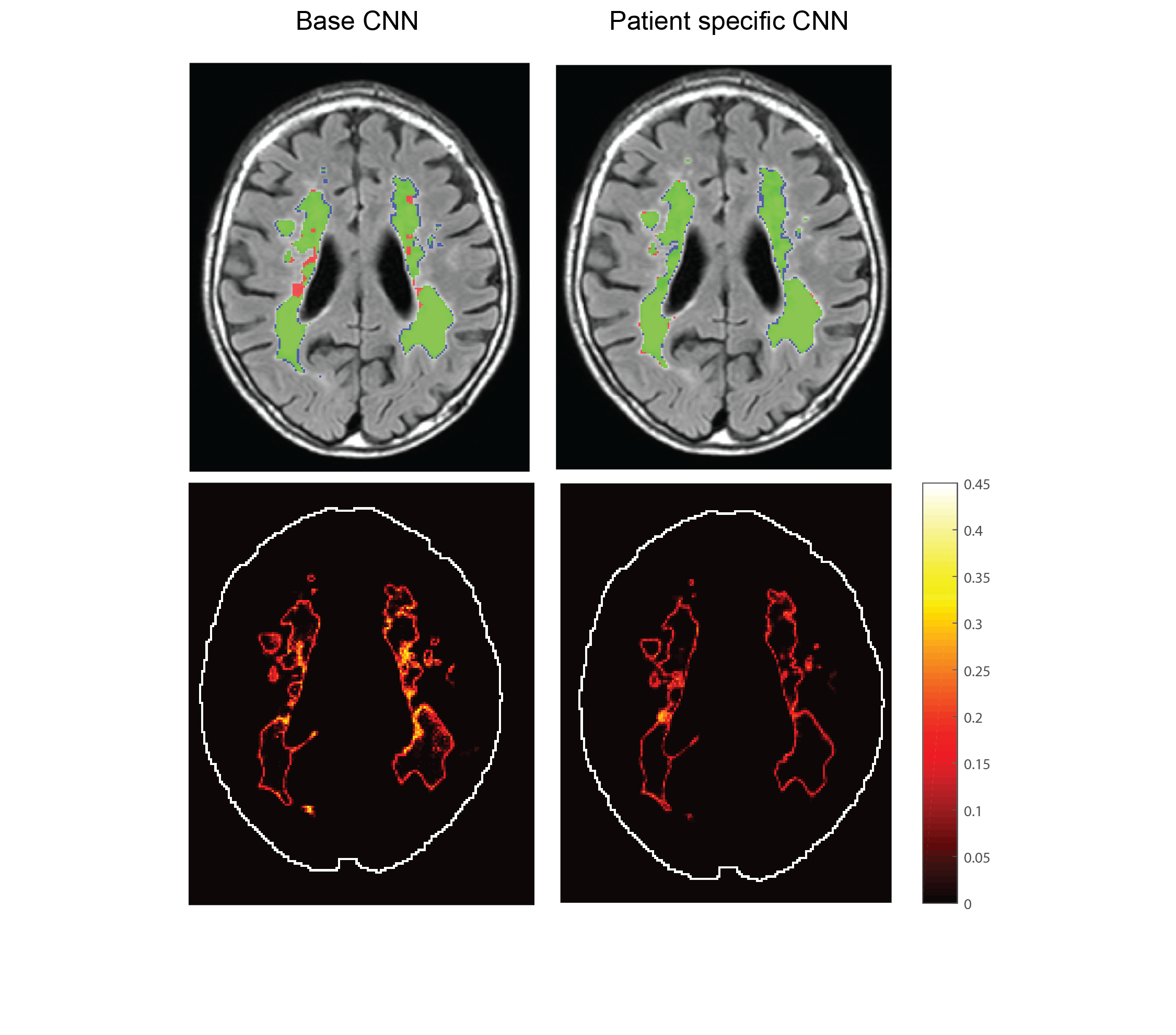}
		\end{tabular}
	\end{center}
	\caption[example]
	{ \label{fig:Fig5} 
		 An example of the uncertainty (SD of softmax probability) of the base CNN and the patient-specific CNN. A high SD means the CNN is uncertain about its decision.}
\end{figure}

\section*{Discussion}
Patient-specific fine-tuning of a CNN with previously acquired MR images yields improvements for quantification of lesions on subsequent MR images. The true positive rate for the detection of liver metastases and the Dice score for the WMH segmentation both increase after patient-specific fine-tuning. Additionally, the certainty of the patient-specific CNN is higher than the certainty of the base CNN.\par
This study shows the potential of the patient-specific fine-tuning step to improve lesion quantification results when follow-up imaging assessment is performed. Previously acquired scans of a patient are proven to hold valuable information to refine a CNN accordingly, resulting in a better performance. The patient-specific fine-tuning step increases the lesion quantification effectiveness for two applications and we expect other applications to benefit from this fine-tuning step as well. \par
The loss function used is not tailored towards the detection of liver metastases but rather to the segmentation of the metastases. In case of large over- or undersegmentation on the baseline scan, it is most cost-effective for the CNN to correct those areas. Large false positive regions are caused by tissue appearances the base CNN is not familiar with, such as radiofrequency ablation (RFA) regions in the liver from earlier treatments. The patient-specific CNN will learn to correctly label the majority of the pixels in these regions. However, after patient-specific fine-tuning pixels with a similar appearance, but of the other class, can be labeled incorrectly. This is an overshoot on the follow-up scan, where the patient-specific CNN corrects its results too much. False negatives on the baseline then might cause some false positives on the follow-up due to a similar appearance and vice versa. As a consequence, more objects are detected incorrectly or more are missed, respectively. This might be overcome by tailoring the loss function towards the detection task, i.e. calculate the loss function on an object base instead of pixel base.\par
The patient-specific fine-tuning method assumes that the baseline and follow-up scan share features that describe the lesions and healthy tissue. However, there can be differences between the baseline scan and the follow-up scan due to (ongoing) treatment and image quality variations. The patients in the brain data set did not undergo any treatment that would be expected to influence WMH features, while most patients in the liver data set did. Therefore, the liver data set shows visually more differences between the consecutive MRI scans than the brain data set. Variation in consecutive images increases the risk of overfitting when the CNN is fine-tuned for many iterations. In addition, fine-tuning for more than 100 iterations did not improve the results compared to fine-tuning for 100 iterations. \par 
Regarding the number of slices to include in the fine-tuning step, both applications yielded the best results when including all slices with lesions present. However, for the WMH segmentation including one or two slices gave similar results. The WMH have similar features throughout the slices and the consecutive MRI exams. This resemblance makes it possible to fine-tune the CNN with only one or two slices, reducing the time and effort invested in annotating the slices. The liver metastases have different features throughout the slices and MRI exams, requiring to fine-tune the CNN with all metastases slices. \par
In addition, an imbalanced data set arises for the liver metastases detection when all the liver slices are included in the fine-tuning step. That is, on average only 30\% of the liver slices contain liver metastases, while 63\% of the brain slices contain WMH. The CNN will then learn to distinguish different types of background instead of the appearance of liver metastases, leading to fewer false positives and unfortunately also fewer true positives. In case of an imbalanced data set, a full lesion annotation is necessary to include only the slices with lesions present and gain improvement with patient-specific fine-tuning. In other cases annotations of a selection of slices might be sufficient to gain improvement on lesion quantification.\par
Moreover, the difference in the percentage of slices containing lesions leads to a considerable difference in the chance that a slice with lesions is selected. This resulted in the selection of slices without liver metastases and the unsuccessful fine-tuning of the CNN. Since brain MRI contain more slices with WMH, the chance to automatically select slices with WMH present is higher, resulting in a successful fine-tuning of the CNN.\par
Weighting the true positives and false positives makes the detection CNN more inclined to label a pixel as lesion. This results in more detected lesions, but also more false positives in comparison with the non-weighted patient-specific CNN. Considering the aim of the detection one might allow more false positive objects in order for more detected lesions. However for the WMH segmentation, the weighting of the false positives, false negatives, and true positive pixels did not result in a better segmentation. \par
The information about the location of the lesions found in the baseline scan could be a valuable information for the detection method. At the moment only the appearances of the lesion and non-lesion tissue are learned. Adding the spatial information of former lesions could aid the method in finding formerly existing lesions. Image registration could be used to detect these lesions in follow-up scans. However such a method should also take into account that more lesions can develop or that lesions can disappear due to treatment. \par

\section*{Conclusions}
In conclusion, patient-specific fine-tuning of a CNN for the quantification of lesions is a viable option to enhance the performance of the method using previously acquired data of the same patient. The CNN is fine-tuned towards the features specific for that patient resulting in a better performance of the CNN. It is important that slices with lesions that represent the features of all lesions are included in the patient-specific fine-tuning step, as well as avoiding imbalanced classes in these slices. In doing so, the patient-specific CNN detected more liver metastases than the base CNN, with the true positive rate increasing from 0.67 to 0.85. The patient-specific CNN segments the WMH better than the base CNN with the Dice score increasing from 0.82 to 0.87.

\begin{acknowledgements}
The authors thank Maarten Niekel, MD, Frank Wessels, MD, and Wouter Veldhuis, MD PhD, for their effort in annotating the liver metastases. \par
This work was financially supported by the project IMPACT (Intelligence based iMprovement of Personalized treatment And Clinical workflow supporT) in the framework of the EU research programme ITEA (Information Technology for European Advancement).

\end{acknowledgements}

\section*{References}
\small{
1.	Litjens, G. et al. A survey on deep learning in medical image analysis. Med. Image Anal. 42, 60–88 (2017).\par
2.	Suzuki, K. Overview of deep learning in medical imaging. Radiol. Phys. Technol. 10, 257–273 (2017).\par
3.	Anwar, S. M. et al. Medical image analysis using convolutional neural networks: A review. J. Med. Syst. 42, 1–13 (2018).\par
4.	Sahiner, B. et al. Deep learning in medical imaging and radiation therapy. Med. Phys. 46, e1–e36 (2019).\par
5.	Mazurowski, M. A., Buda, M., Saha, A. and Bashir, M. R. Deep learning in radiology: An overview of the concepts and a survey of the state of the art with focus on MRI. J. Magn. Reson. Imaging 49, 939–954 (2018).\par
6.	Dhara, A. K. et al. Segmentation of Post-operative Glioblastoma in MRI by U-Net with Patient-Specific Interactive Refinement. in Crimi A., Bakas S., Kuijf H., Keyvan F., Reyes M., van Walsum T. (eds) Brainlesion: Glioma, Multiple Sclerosis, Stroke and Traumatic Brain Injuries. BrainLes 2018. Lecture Notes in Computer Science 11383, 115–122 (Springer, Cham, 2019).\par
7.	Laves, M., Bicker, J., Kahrs, L. A. and Ortmaier, T. A dataset of laryngeal endoscopic images with comparative study on convolution neural network-based semantic segmentation. Int. J. Comput. Assist. Radiol. Surg. 14, 483–492 (2019).\par
8.	Hermessi, H., Mourali, O. and Zagrouba, E. Transfer learning with multiple convolutional neural networks for soft tissue sarcoma MRI classification. in Proc. SPIE 11041, Eleventh International Conference on Machine Vision (ICMV 2018) (2018). doi:10.1117/12.2522765\par
9.	Tajbakhsh, N. et al. Convolutional neural networks for medical image analysis: Full training or fine tuning? IEEE Trans. Med. Imaging 35, 1299–1312 (2016).\par
10.	Wang, G. et al. Interactive medical image segmentation using deep learning with image-specific fine-tuning. IEEE Trans. Med. Imaging 37, 1562–1573 (2018).\par
11.	Bredell, G., Tanner, C. and Konukoglu, E. Iterative Interaction Training for Segmentation Editing Networks. in Shi Y., Suk HI., Liu M. (eds) Machine Learning in Medical Imaging. MLMI 2018. Lecture Notes in Computer Science, vol 11046 363–370 (Springer International Publishing, 2018). doi:10.1007/978-3-030-00919-9\par
12.	Jansen, M. J. A. et al. Evaluation of motion correction for clinical dynamic contrast enhanced MRI of the liver. Phys. Med. Biol. 62, 7556–7568 (2017).\par
13.	Jansen, M. J. A. et al. Liver segmentation and metastases detection in MR images using convolutional neural networks. J. of Medical Imaging, 6(4), 044003 (2019).\par
14.	Aalten, P. et al. The Dutch Parelsnoer Institute - Neurodegenerative diseases; methods, design and baseline results. BMC Neurol. 14, 1–8 (2014).\par
15.	Kuijf, H. J. et al. Standardized Assessment of Automatic Segmentation of White Matter Hyperintensities; Results of the WMH Segmentation Challenge. IEEE Trans. Med. Imaging 99, 1–1 (2019).\par
16.	Boomsma, M. J. F. et al. Vascular cognitive impairment in a memory clinic population: Rationale and design of the “Utrecht-Amsterdam Clinical Features and Prognosis in Vascular Cognitive Impairment” ( TRACE-VCI ) Study. JMIR Res. Protoc. 6, (2017).\par
17.	Wardlaw, J. M. et al. Neuroimaging standards for research into small vessel disease and its contribution to ageing and neurodegeneration. Lancet Neurol. 12, 822–838 (2013).\par
18.	Penny, W., Friston, K., Ashburner, J., Kiebel, S. and Nichols, T. Statistical parametric mapping: The analysis of functional brain images. (Academic Press, 2007).\par
19.	Yu, F. and Koltun, V. Multi-Scale Context Aggregation by Dilated Convolutions. in ICLR (2016). doi:10.16373/j.cnki.ahr.150049\par
20.	Prins, N. D. and Scheltens, P. White matter hyperintensities, cognitive impairment and dementia: an update. Nat. Rev. Neurol. 11, 157–165 (2015).\par
21.	Au, R. et al. Association of White Matter Hyperintensity Volume With Decreased Cognitive Functioning. Arch Neurol. 63, 246–250 (2006).\par
22.	Wolters, F. J. et al. Cerebral perfusion and the risk of dementia. Circulation 136, 719–728 (2017).\par
23.	Haley, A. P. et al. Subjective cognitive complaints relate to white matter hyperintensities and future cognitive decline in patients with cardiovascular disease. Am. J. Geriatr. Psychiatry 17, 976–985 (2009).\par
24.	Sander, J., de Vos, B. D., Wolterink, J. M. and Isgum, I. Towards increased trustworthiness of deep learning segmentation methods on cardiac MRI. in Proc. SPIE 10949, Medical Imaging 2019: Image Processing 1094919 (2019). doi:10.1117/12.2511699\par
25.	Gal, Y. and Ghahramani, Z. Dropout as a Bayesian approximation: Representing model uncertainty in deep learning. in International Conference on Machine Learning (ICML) 2016, 1050–1059 (2016).

}

\end{document}